\documentstyle[psfig]{mn}

\newif\ifAMStwofonts  
  
\newcommand{\Ha}{H$\alpha$}

\newcommand{\Msolar}{M$_{\odot}$}                           

\ifoldfss  
  \ifCUPmtlplainloaded \else  
    \NewTextAlphabet{textbfit} {cmbxti10} {}  
    \NewTextAlphabet{textbfss} {cmssbx10} {}  
    \NewMathAlphabet{mathbfit} {cmbxti10} {} 
    \NewMathAlphabet{mathbfss} {cmssbx10} {} 
  \fi  
  \ifAMStwofonts  
    \ifCUPmtlplainloaded \else  
      \NewSymbolFont{upmath} {eurm10}  
      \NewSymbolFont{AMSa} {msam10}  
      \NewMathSymbol{\upi}     {0}{upmath}{19}  
      \NewMathSymbol{\umu}     {0}{upmath}{16}  
      \NewMathSymbol{\upartial}{0}{upmath}{40}  
      \NewMathSymbol{\leqslant}{3}{AMSa}{36}  
      \NewMathSymbol{\geqslant}{3}{AMSa}{3E}

       \let\ge=\geqslant  
    \fi  
  \fi  
\fi 
  
\ifnfssone  
  \newmathalphabet{\mathit}  
  \addtoversion{normal}{\mathit}{cmr}{m}{it}  
  \addtoversion{bold}{\mathit}{cmr}{bx}{it}  
  \newmathalphabet{\mathbfit} 
  \addtoversion{normal}{\mathbfit}{cmr}{bx}{it}  
  \addtoversion{bold}{\mathbfit}{cmr}{bx}{it}  
  \newmathalphabet{\mathbfss} 
  \addtoversion{normal}{\mathbfss}{cmss}{bx}{n}  
  \addtoversion{bold}{\mathbfss}{cmss}{bx}{n}  
  \ifAMStwofonts  
    \ifCUPmtlplainloaded \else  
      \UseAMStwoboldmath  
      \makeatletter  
      \new@mathgroup\upmath@group  
      \define@mathgroup\mv@normal\upmath@group{eur}{m}{n}  
      \define@mathgroup\mv@bold\upmath@group{eur}{b}{n}  
      \edef\UPM{\hexnumber\upmath@group}  
      \new@mathgroup\amsa@group  
      \define@mathgroup\mv@normal\amsa@group{msa}{m}{n}  
      \define@mathgroup\mv@bold\amsa@group{msa}{m}{n}  
      \edef\AMSa{\hexnumber\amsa@group}  
      \makeatother  
      \mathchardef\upi="0\UPM19  
      \mathchardef\umu="0\UPM16  
      \mathchardef\upartial="0\UPM40  
      \mathchardef\leqslant="3\AMSa36  
      \mathchardef\geqslant="3\AMSa3E  

       \let\ge=\geqslant  
    \fi  
  \fi  
\fi 
  
\ifnfsstwo  
  \DeclareMathAlphabet{\mathbfit}{OT1}{cmr}{bx}{it}  
  \SetMathAlphabet\mathbfit{bold}{OT1}{cmr}{bx}{it}  
  \DeclareMathAlphabet{\mathbfss}{OT1}{cmss}{bx}{n}  
  \SetMathAlphabet\mathbfss{bold}{OT1}{cmss}{bx}{n}  
  \ifAMStwofonts  
    \ifCUPmtlplainloaded \else  
      \DeclareSymbolFont{UPM}{U}{eur}{m}{n}  
      \SetSymbolFont{UPM}{bold}{U}{eur}{b}{n}  
      \DeclareSymbolFont{AMSa}{U}{msa}{m}{n}  
      \DeclareMathSymbol{\upi}{0}{UPM}{"19}  
      \DeclareMathSymbol{\umu}{0}{UPM}{"16}  
      \DeclareMathSymbol{\upartial}{0}{UPM}{"40}  
      \DeclareMathSymbol{\leqslant}{3}{AMSa}{"36}  
      \DeclareMathSymbol{\geqslant}{3}{AMSa}{\"3E}  

       \let\ge=\geqslant  
    \fi  
  \fi  
\fi 
  
\ifCUPmtlplainloaded \else  
  \ifAMStwofonts \else 
    \def\upi{\pi}  
    \def\umu{\mu}  
    \def\upartial{\partial}  
  \fi  
\fi  
  
\title[Sub-arcsecond Radio Observations of NGC\,3077]
{Sub-arcsecond Radio Observations of the Dwarf Starburst Galaxy NGC\,3077 }  
  
\author[D. Rosa--Gonz\'alez]  
{Daniel Rosa--Gonz\'alez$^{1,2}$ \\
$^1$ INAOE, Luis Enrique Erro 1. Tonantzintla, Puebla 72840. M\'exico. 
\\ $^2$  Astrophysics Group, Blackett Laboratory, Prince Consort Road, London SW7 2AZ, United Kingdom. 
}
\date{Accepted  .  
      Received ;  
      in original form \today\  (Version 006)}  
  
\pagerange{\pageref{firstpage}--\pageref{lastpage}}  
\pubyear{2003}  
  
\begin{document}  
  
\maketitle  
  
\label{firstpage}  
  
\begin{abstract}  
We present the first sub-arcsecond  radio observations of
the  nearby dwarf  starburst galaxy  NGC~3077 obtained with the  MERLIN interferometer. 
We have detected two resolved sources which are coincident with 
the positions of two discrete X-ray sources  detected by Chandra.
One of the radio sources is associated with a supernova remnant and the 
observed radio flux is consistent with having a non-thermal origin. 
The age of the SNRs of about 760 years is between  
the average age of the SNRs detected in M82 and 
those detected in the Milky Way and the Large Magellanic Cloud. 
We use this detection to calculate a star formation
rate (SFR) of 0.28 \Msolar year$^{-1}$ which is similar to the SFR calculated
by using  far infrared and millimeter observations but larger than
the SFR given by optical recombination lines corrected for extinction.
The other compact radio source detected by  MERLIN which is coincident 
with the position of an X-ray binary,  has the properties of 
an HII region with a flux density of about 747 $\mu$Jy which corresponds to an  
ionizing flux of 6.8$\times 10^{50}$s$^{-1}$. A young  massive stellar cluster 
with a mass of $\sim$ 2$\times 10^5$\Msolar\, detected by 
the {\it Hubble Space Telescope} could be the responsible for the production
of the ionizing flux.
\end{abstract}  
  
\section{Introduction}  
 
The  study of  small  starburst  galaxies such  as  NGC~3077 has  been
favoured  recently  mainly  because  of the  implications  that  these
objects have  in the  {\it standard} model  of galaxy  evolution (e.g.
Baugh,  Cole \& Frenk  1996).  In  the hierarchical  scenario, smaller
systems form first and then  become the building blocks of the massive
galaxies that are observed in the local universe.  These systems 
which are the most numerous type of galaxies,  could
be responsible  for an important  fraction of the reionization  of the
universe.  Moreover,  due to the low gravitational potential 
of those galaxies  the interstellar medium 
might be allowed to escape from the host more easily
contributing  to the enrichment  of the intergalactic medium at early epochs.
However the expelling of newly processed matter depends  not only on the
mass of the host and the power of the burst but also on the distribution of 
the interstellar medium and the presence of a  dark matter 
halo surrounding the host  galaxy (e.g. Silich \& Tenorio-Tagle 2001).

Nearby compact starburst galaxies  are excellent laboratories in which
to study the starburst phenomenon. In fact, compact starburst galaxies
have been used to test the validity of different star forming tracers
(e.g.   Rosa-Gonz\'alez, Terlevich  \& Terlevich  2002); to  study the
interaction    of   the   burst    with   the    interstellar   medium
(e.g. Martin~1998;  Silich ~et~al.~2002); and to  study the enrichment
of the intergalactic medium due to the break out of superbubbles (e.g.
Kunth  et al.   2002). 

It  is  only  in  the  nearby universe  that  the  physical  processes
related to the current starburst event can be studied in great
detail.  The  presence   of  a   recent  starburst  event---triggered   
by  the interaction  of NGC~3077 with  M~81 and  M~82---has been  confirmed by
several independent tracers.  The IUE ultraviolet (UV) spectra revealed the
presence  of   massive  stars  not   older  than  7$\times$10$^7$years
(Benacchio \& Galletta 1981).  
The  peak of the  P$\alpha$  nebula -- a tracer of  young star formation  regions -- 
(Meier, Turner  and Beck  2001;  B\"oker et  al.  1999,  Figure~\ref{PaImage})
is located between two  CO complexes detected by 
the Owens Valley Radio Observatory (Walter et al. 2002).
Walter et al.  conducted a comprehensive multiwavelength study
of NGC~3077, relating  the atomic and molecular gas  with the observed
HII regions.  By combining CO and emission line observations, they concluded that  
the star formation efficiency -- defined as the ratio between the \Ha\, 
luminosity  and the total amount of molecular gas  -- in NGC~3077 is 
higher than the corresponding value in M~82. They conclude that 
the recent star formation activity in NGC~3077 and M~82 
is probably due to their interaction with M~81.

The extinction corrected  H$\alpha$  flux  indicates a star formation  rate (SFR)
of about  0.05  M$_\odot$yr$^{-1}$, concentrated  in  a  region of  about
150~pc in diameter (Walter et al. 2002). 
This value is lower than the SFR given by extinction--free tracers of the SFR 
like the mm continuum or the far infrared  (FIR). In fact, the SFR given by observations at 2.6 mm by 
Meier~et~al.~(2001) gives a SFR of  0.3 in agreement with 
FIR  measurements from Thronson, Wilton \&  Ksir (1991).


Maps of radio emission can reveal SNRs and HII regions 
as well as other tracers of  the star forming activity within a galaxy,
such as recombination emission lines  and FIR radiation.   
They are closely related to
the  evolution of  massive stars  also,  therefore,  to the
recent star forming history of the galaxy (e.g.  Muxlow et al. 1994).
In this paper we present for the first time radio maps of NGC~3077 with 
sub-arcsecond  angular resolution.

For consistency with previous publications of radio observations 
we assume a distance to NGC~3077
 of 3.2 Mpc  throughout the paper (e.g. Tammann \& Sandage 1968).

\section{Observations}
\begin{figure}
\setlength{\unitlength}{1cm}           
\begin{picture}(7,8.0)         
\put(-0.5,-0.){\includegraphics{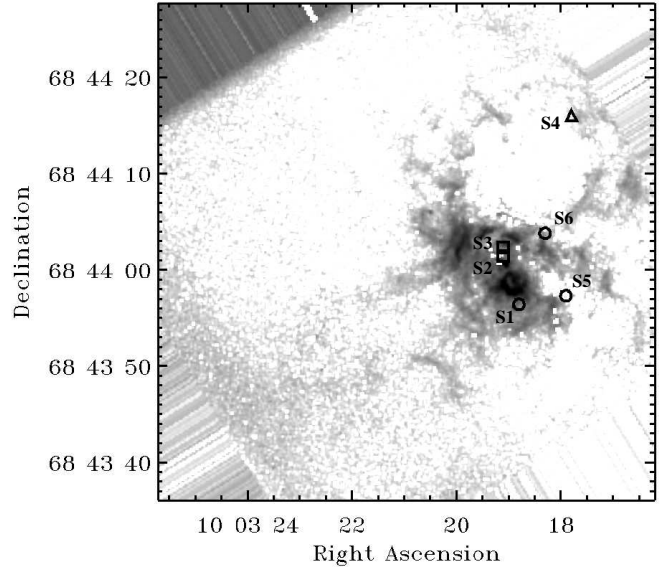}}
\end{picture}
\caption{\label{PaImage} P$\alpha$ image of the central part of NGC~3077 
showing the regions of recent star
formation activity (B\"oker et al. 1999).
 The  symbols are the discrete
sources detected by  Chandra. The circles and squares 
represent possible SNRs and accreting objects, respectively.
The triangle indicates the supersoft source characterized in the X-ray by the
absence of emission above 0.8 keV. The observed FOV of 30\arcsec\, in radius
cover the whole P$\alpha$ emission represented in this figure.}
\end{figure}

NGC~3077 was observed in May 2004 
using the MERLIN interferometer, including the Lovell telescope at Jodrell Bank. 
NGC~3077 and the phase reference source I0954+658 were observed 
during a total time of about 21.5 hours. 
The flux density scale was calibrated 
assuming a 7.086 Jy for 3C~286. The observations were made using the
wide field mode and  an  observing frequency of 4.994~GHz 
in each hand of circular polarization using a bandwidth of 13.5~MHz.
Visibilities corrupted by instrumental phase errors, telescope errors,
or external interference were flagged and discarded by using the local
software provided by   Jodrell Bank Observatory. 
The unaberrated field of view of 30\arcsec\, in radius allows us to cover 
the main active star forming region revealed by  emission line images (see
Figure~\ref{PaImage}).  
The  data were naturally weighted using a cell size of 0.015\arcsec.
The images were deconvolved using the {\small CLEAN} algorithm 
described by H{\" o}gbom (1974).
The rms noise over source-free areas of the image was $\sim$60
$\mu$Jy~beam$^{-1}$. 
The final MERLIN spatial resolution after restoring the image with  a circular
Gaussian beam was  0.14\arcsec. At the assumed distance of  NGC~3077 
this angular size corresponds roughly to 2 pc. 

\section{\large \bf Discrete X-ray sources and their radio counterparts}

At X-ray wavelengths,  a star formation event is  characterized by the
presence of diffuse emission associated  with hot gas 
but also by the presence of compact objects associated with SNRs and  
high mass X-ray binaries  (e.g. Fabbiano  1989).  
Recent  Chandra observations of NGC~3077 
(Ott, Martin \& Walter 2003)
revealed  the presence  of  hot gas  within expanding  H$\alpha$
bubbles.  Ott  et~al.  found  that the  rate at which  the hot  gas is
deposited into the halo is a  few times the SFR measured by Walter et
al. (2002).  
Ott et.~al (2003) found 6 discrete X-ray sources
close to the centre of the galaxy, but also associated with  
bright HII regions (see Figure~\ref{PaImage}). The details of these
are given in Table~\ref{tab:points}.

\begin{table*}\begin{center}
\caption{\small Discrete X-ray sources detected in NGC~3077.  Sources marked
  with an asterisk have been detected at 5 GHz by  the present observations.
The proposed type is based only in the X-ray observations. X-ray unabsorbed fluxes and
  luminosities (columns 6 and 7) are based on the best fitted spectra for each
  individual source (see text for details).} 
\label{tab:points}
\begin{tabular}{lllclcc}\\
\hline\hline 
Source & RA (J2000) & DEC (J2000) & X-Ray Photons & Proposed Type &
Flux & Luminosity\\
  \  &   \  & \  &  (counts) & \  & ($\times 10^{-15}$ erg cm$^{-2}$ s$^{-1}$) & ($\times 10^{37}$ erg s$^{-1}$)    \\
\hline
S1$^\ast$ & 10~03~18.8 & +68~43~56.4 &  133$\pm$12 & SNR      & 96.58 & 14.98  \\ 
S2 & 10~03~19.1 & +68~44~01.4 & 114$\pm$11 & Accreting        & 71.56 & 11.10  \\ 
S3$^\ast$ & 10~03~19.1 & +68~44~02.3 & 119$\pm$11 & Accreting & 65.14 & 10.10  \\ 
S4 & 10~03~17.8 & +68~44~16.0 & 37$\pm$7 & Supersoft Source   & 5.92  &  0.92  \\ 
S5 & 10~03~17.9  & +68~43~57.3 & 17$\pm$4 & SNR               & 2.06  &  0.32  \\
S6 & 10~03~18.3 & +68~44~03.8 & 17$\pm$4 & SNR                & 17.09 &  2.65  \\
\hline
\end{tabular}\end{center}
\end{table*}

The X-ray  spectrum of S1, S5  and S6 was  found to peak in  the range
$\sim$0.8--1.2~keV. 
The X-ray spectral properties suggest that S1, S5 and S6 
are SNRs. The X-ray spectra of these sources was best fitted with a 
Raymond-Smith collisional plasma (Ott  et.~al 2003). The obtained fluxes and
luminosities are given in Table~\ref{tab:points}. 
Ott  and collaborators derived a radio  continuum spectral index
of $\alpha=-$0.48  for S1.  However,  this determination was  based on
single dish  radio observations with  a resolution of 69\arcsec\,  (Niklas et
al.   1999) and  VLA observations  with a  resolution of  54\arcsec\,  (Condon
1987).  Recent  VLA observations of NGC~3077  at 1.4 GHz  
have reported an unresolved source located in the same position 
(Walter et al. 2002). 

\begin{figure}
\setlength{\unitlength}{1cm}           
\begin{picture}(7,7.)         
\put(0.0,-1){\includegraphics{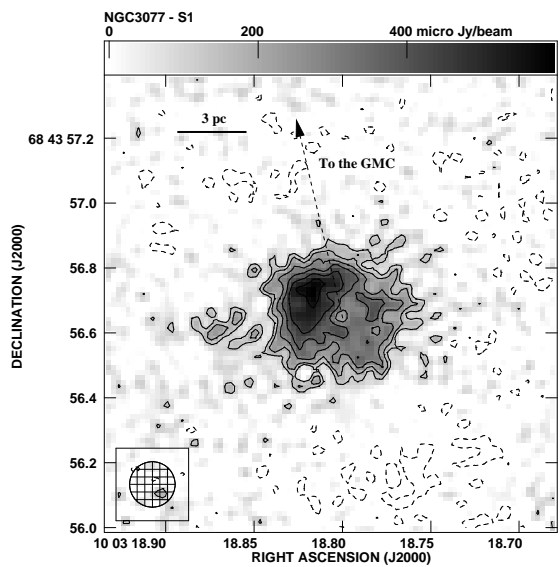}}
\end{picture}
\caption{\label{Radio_S1} 
MERLIN 5 GHz radio map of the supernova remnant coincident with
the X-ray source S1. The obtained rms was 80 $\mu$Jy  beam$^{-1}$. 
The contours are at -2, 2, 3, 4.5, 6, 7.5, 9  times the  rms.  
 The maximum flux  within the  image is
595\,$\mu$Jy  beam$^{-1}$. The restoring beam of 0.14\arcsec\, in
diameter is plotted in the bottom right corner of the image.
The dashed line points to the position of the GMC described in the text.}
\end{figure}
 
In  Figure~\ref{Radio_S1} we present the resultant MERLIN radio map 
which shows strong emission coincident, within the Chandra positional errors,  
with the position of the S1 X-ray  source. 

The semi--circular morphology of the source showing the presence of 
bright knots is similar to  SNRs observed in other galaxies (e.g. 
43.18+58 3 in M~82, Muxlow et al. 1994 or SN1986J in NGC 891, P\'erez--Torres, 
Alberdi \& Marcaide 2004). 
The SNR has a noticeable asymmetry that could be due to  interaction  
with the surrounding media.
In fact, a giant molecular cloud (GMC) with a mass of $\sim$ 10$^7$\Msolar\,  
has been detected in CO (Meier et al. 2001).  This GMC has a projected size of 
79$\times$62 parsecs  (5.3\arcsec$\times$4.1\arcsec) and the centroid of the
emission is localized just 1.1\arcsec\,  from the SNR in the north east direction   
\footnote{
The position of the GMC is: 10$^{\rm h}$:03$^{\rm m}$:18.85$^{\rm s}$,
+68$^{\rm o}$:43\arcmin:57.8\arcsec}.  
The interaction of the remmant with this giant molecular cloud 
could be the cause of the asymmetry detected in the radio map. 

Due to the non-thermal nature  of the SNR emission, radio sources with
temperatures higher  than 10$^4$K  can be unambiguously  identified as
SNRs. For this radio source we measure a peak flux of 595 $\pm 60 \mu$Jy/beam which
corresponds to a brightness 
temperature of 1.24$\pm 0.125\times 10^4$K for the given MERLIN spatial 
resolution at 5~GHz. The angular size of the source 
-- measured on the map using the 3$\times$ rms contour -- is roughly
0.5\arcsec\,   
which corresponds to a physical size of about 8~pc at the assumed distance of
NGC~3077. If the SNR has expanded  with a constant velocity of 
5000~km~s$^{-1}$ (e.g. Raymond 1984) we deduce that the 
progenitor star exploded about 760 years ago.  
This age is longer than the average age of the SNRs detected in M~82 ($\sim$200 years)
and shorter than the average ages of  2000 and 3000 years  of the  SNRs detected in the Milky
Way and in the LMC respectively (Muxlow et al. 1994). 

The assumption of constant expansion velocity is quite naive, and the size
of the  observed supernova remnant depends,  among others,  on the density of the 
interstellar medium, initial kinetic energy or the size of the cavity created 
by the stellar wind prior to the supernova explosion.
However, there is observational evidence of the existence of cavities 
created by the stelar winds of the supernova progenitor.   
In these cavities with densities as low as 0.01 cm$^{-3}$ the 
velocity of the SNR can reach values of 5000 km~s$^{-1}$
(e.g. Tenorio-Tagle et al. 1990, 1991 and references therein). 
If that is the case of the SNR observed in NGC~3077 the assumed expansion velocity 
of the supernova blast and the estimated age are reliable.
In any case we use the value of  5000 km~s$^{-1}$ for the expansion 
velocity in order to compare our results with those 
from observations of M~82 (Muxlow et al. 1994).

The life time of massive star progenitors of the SNRs observed in NGC~3077
is much shorter than the Hubble time, therefore the number of observed 
remnants can be used as a tracer of the current star formation rate. 
The fact that we observed only one SNR with an estimated age of 
760 years can be converted into a supernova rate ($\nu_{\rm SN}$) of 1.3$\times 10^{-3}$ year$^{-1}$.
The supernova rate can be translated to the SFR by using,

\begin{equation}\label{SFR}
 \rm SFR (M > 5\,M_\odot) = 24.4 \times  \left(\frac{\nu_{\rm SN}}{year^{-1}} \right)   \rm M_\odot year^{-1}
\end{equation}

Equation~\ref{SFR} was calculated by using a Scalo IMF, with a lower mass  limit
of 5\,M$_\odot$ and an upper limit of 100\,M$_\odot$ (Condon 1992). 
In any galaxy most of the stellar mass is located in low mass stars, 
therefore to calculate the total SFR, including the mass contained
in stars with masses lower than 5\,M$_\odot$ we need to multiply the 
factor 24.4 in Equation~\ref{SFR} by 9. 
Combining Equation~\ref{SFR} with the calculated supernova rate we obtain a  SFR for NGC~3077 of 
0.28 M$_\odot$ year$^{-1}$. 
The estimated SFR is in reality a upper limit because the presented observations
are sensitive to older  SNRs which we did not detect.
Assuming that the flux of a SNR decay with a rate of $\sim$1\%~year$^{-1}$ (e.g. Kronberg \& Sramek
1992, Muxlow et al. 1994), the three sigma detection limit of 180~$\mu$Jy/beam
allows to detect older SNRs with ages of 880 years, implying a lower supernova rate
$\nu_{\rm SN}$= 1.14 $\times 10^{-3}$ year$^{-1}$.
This fact produce a change in the calculated SFR of about 14\%.

The flux density of the SNR was calculated by using the AIPS task
{\small IMSTAT} which add the observed fluxes  within the area defined by the
SNR at three times the noise level. 
We obtain a flux of  2100$\pm 175$ $\mu$Jy. 
For this SNR, the relation between the size and the flux density is consistent with the
relation found by Muxlow et al. (1994) for a sample of SNRs detected in M~82
and the LMC. This relation, where the flux density is inversely proportional to 
the diameter, is not consistent with simple adiabatic losses in a
synchrotron-emitting source and an extra source of relativistic particles must
come from other reservoir of energy in the  form of thermal or kinetic 
energy present in the remnant (Miley 1980).

The ratio between radio and  X-ray fluxes R$_{\rm r-x}$=~5$\times~10^9$ F$_{\rm 5GHz}$/F$_{\rm x}$
is  highly variable and depends on  the nature of the object, 
the surrounding media prior to the supernova
explosion and the time at which the supernova remmant is observed.
Table~\ref{Tab:RadioXray} shows the  value of R$_{\rm r-x}$ for a small sample of
SNRs which include the  brightest SNRs in our galaxy, Cassiopeia A and Crab nebula.   
For the galactic SNRs the radio data which includes morphological type, 
flux and spectral index, was obtained from the Green catalogue (Green 2004).
This catalogue is based on observations at 1 GHz. We  used   the given 
spectral index to estimate the flux at 5 GHz in order to compare with our observations.
For the case of SN1988Z we used the data compiled by Aretxaga et al. (1999) 
and for the case of  NGC7793-S26,  the data from Pannuti et al. (2002).
The X-ray data are from the compilation  by Seward et al. (2005) except 
S1 and S3 from Ott  et al. (2003), SN1006AD from Dyer et al. (2001),  SN1988Z from Aretxaga et
al. (1999) and NGC7793-S26  from Pannuti et al. (2002).
The value of R$_{\rm r-x}$ goes from 0.02$\times 10^{-4}$ for the case of
SN1006AD to  0.2722 for Vela. 
The value  of  R$_{\rm r-x}$ for S1 is within the observed range.

The other candidates to SNRs -- S5 and S6 -- were not detected by the present
observations. The Xray fluxes of S5 and S6 are 50 and 6 times  respectively 
lower than the Xray flux of S1 (see Table~\ref{tab:points}). 
Assuming that in both cases the ratio between the X-ray and radio fluxes is
equal to the ratio observed in S1, R$_{\rm r-x}$(S1)= 10.8$\times 10^{-4}$ then the expected
radio flux for S5 and S6 is below the 3$\sigma$ detection limit. 

\begin{table*}
\begin{center}
\caption{\label{Tab:RadioXray} 
Observed properties of a small smaple of SNRs. 
Second column are the type of the galactic SNRs based on radio observations.  
Types S and F correspond to shell or filled-centre structure respectivelly and 
type C if the SNR shows a composite morphology. Third column are the radio spectral index.
Radio and X-ray fluxes are given in the fourth and fifth column. 
The last column shows the ratio between radio and X-ray fluxes  as defined in the text.
}
\begin{tabular}{lccccr}\hline
 Name          & Type & Spectral Index   &   Flux at 5 GHz & X-ray Flux &  R$_{\rm r-x}$ \\ 
             \ &  &\    &   (mJy)         & ($\times 10^{-15}$erg s$^{-1}$cm$^{-2}$) & ($\times$10$^{-4}$) \\  \hline 
              S1  & --  &   --   &   2.10e+00  &   9.66e+01  &    10.87 \\
  Cassiopeia   A  & S   &  0.77  &   7.88e+05  &   2.06e+07  &    19.12 \\
           Tycho  & S   &  0.61  &   2.10e+04  &   1.99e+06  &     5.27 \\
          Kepler  & S   &  0.64  &   6.78e+03  &   6.85e+05  &     4.95 \\
            W49B  & S   &  0.48  &   1.76e+04  &   9.00e+06  &     0.98 \\
          RCW103  & S   &  0.50  &   1.25e+04  &   1.70e+07  &     0.37 \\
       SN1006-AD  & S   &  0.60  &   7.23e+00  &   2.00e+05  &     0.02 \\
          SN1181  & F   &  0.10  &   2.81e+04  &   2.70e+04  &     520  \\
            Crab  & F   &  0.30  &   6.42e+05  &   2.88e+07  &    11.14 \\
            Vela  & C   &  0.60  &   6.66e+05  &   8.87e+04  &     3755 \\
      G292.0+1.8  & C   &  0.40  &   7.88e+03  &   2.09e+06  &     1.89 \\
SN1988Z           & --  &   --   &   5.30e-01  &   4.00e+01  &     6.62 \\
     NGC7793-S26  & --  &  0.60  &   1.24e+00  &   3.90e+01  &    15.90 \\ \hline 
\end{tabular}
\end{center}
\end{table*}

\begin{figure}
\setlength{\unitlength}{1cm}           
\begin{picture}(7,7.)         
\put(0.,-1){\includegraphics{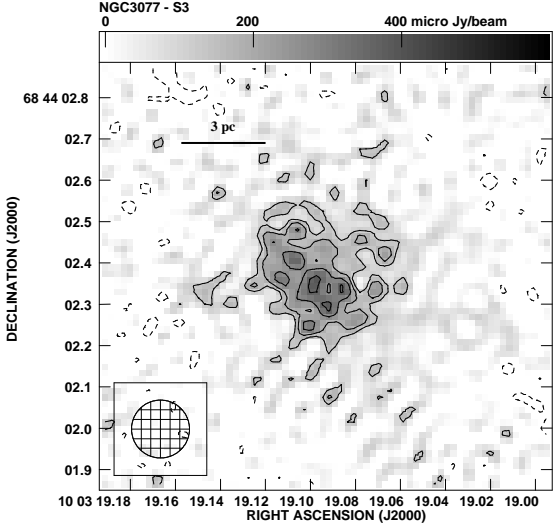}}
\end{picture}
\caption{\label{Radio_S3} MERLIN 5 GHz radio map of the source coincident with
the compact X-ray source S3. Contours and restoring beam size as in Figure~\ref{Radio_S3}. 
The maximum flux  within the  image is 390\, $\mu$Jy beam$^{-1}$.
}
\end{figure}

In the Chandra image of NGC~3077,  the sources S2 and S3 are separated
by   1\arcsec.    This   separation    is   roughly   the   FWHM   of   the
point-spread-function.   S2 and S3  each have  an  X-ray spectrum that was
fitted to a power law.
These objects were  proposed to be either X-ray  binaries or background
active galactic nuclei  (Ott et al. 2003).
Figure~\ref{Radio_S3} shows the map of the radio source coincident with 
the position of S3. The measured intensity peak is 390$\pm$ 60 $\mu$Jy beam$^{-1}$
which is equivalent to a brightness temperature of about 8000 K,  
typical of an HII region. 

Notice that these observations did not rule out 
the possibility that the observed radio source is an old SNR
and further observations at longer wavelength with similar angular resolution and 
sensitivity are necessary. S3 has a R$_{\rm r-x}$= 5.73$\times 10^{-4}$ 
which is within the range of values observed in SNRs. 
However the observed ratio R$_{\rm r-x}$  in well studied SNRs
(Table~\ref{Tab:RadioXray}) covers at least  5 order of magnitudes 
therefore the  R$_{\rm r-x}$ value can not be used to 
discriminate between SNRs and other kind of objects.

Due to the calculated brightness temperature and 
the X-ray spectrum of S3 the probability that the observed radio emission is
coming from an HII region is the most plausible. 
S3 is probably embebed by the HII nebula but the Xray flux  from the binary and
the radio flux have not a common origin.

\begin{figure}
\setlength{\unitlength}{1cm}           
\begin{picture}(7,7)         
\put(-1.,-2.){\includegraphics{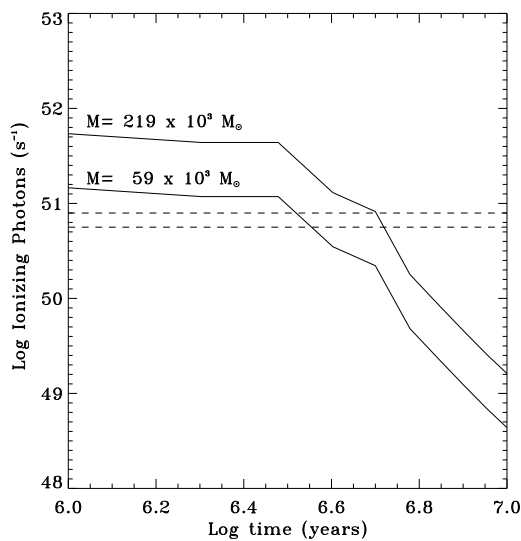}}
\end{picture}
\caption{\label{HarrisCluster} Evolution of the production rate of ionizing photons
from the SB99 code. The solid lines are the result for two different cluster
masses and the dashed horizontal lines are the estimated number of ionizing photons based
on the radio observations.}
\end{figure}

The free--free emission associated to an HII region
can be translated to the number of ionizing photons, $\rm N_{\rm UV}$ by (Condon 1992), 

\begin{equation}\label{eq:Nuv}
\rm N_{\rm UV} \ge 6.3 \times \left(\frac{T_e}{10^4 \rm K}\right)^{-0.45}
\left(\frac{\nu}{\rm GHz}\right)^{0.1}\left(\frac{L_T}{10^{20} \rm W Hz^{-1}}\right)
\end{equation}
The  flux density estimated for the HII region associated with S3 was 747$\pm$127 $\mu$Jy.  
If we assume that the observed flux has a thermal origin, we 
can estimate the thermal luminosity, L$_{\rm T}$.  
We calculated that the number of UV photons coming from the 
observed region is 6.77$\pm$1.15$\times 10^{50}$s$^{-1}$.   
Taking into account that  O stars produce between 2$\times 10^{49} $s$^{-1}$
and 1$\times 10^{50} $s$^{-1}$  UV photons (e.g. Mas-Hesse \& Kunth 1991) we
conclude that only a few 
massive stars are responsible for the ionization of the observed nebula. 
A bright stellar cluster (named cluster \#~1 by Harris et al. 2004) 
was detected by the {\it Hubble Space Telescope} at just 0.3\arcsec\, from  the detected HII region. 
This cluster  with a mass between 59$\times$ 10$^3$ and 219$\times$ 10$^3$
\Msolar\, and an estimated age of 8 Myr  could be the source of ionization of the observed
HII nebula.  We  used  the SB99 synthesis model
(Leitherer et al. 1999) to estimate the evolution of the ionizing photons for
the case of a Salpeter IMF and masses varying between 0.1 and 100\Msolar. 
Figure~\ref{HarrisCluster} shows the results of the SB99 code for a young cluster
with masses within the ranges of masses of the  \#1 cluster.
Using the calculated number of ionizing photons from equation~\ref{eq:Nuv} we
estimate that the HII region has an age  between 3.3 and 5.3 Myr  which
is  just 2 times lower than the age  obtained by Harris and collaborators
based on optical observations. 
 
Notice that the stellar cluster associated with the HII region 
is the most massive young cluster observed in NGC~3077 and deeper observations
are necessary to obtain radio images of the HII regions associated  to 
the other 55 clusters detected in NGC~3077 by the {\it Hubble Space Telescope}.




One of the most  interesting discrete X-ray  source in NGC~3077 is  S4.  This
source exhibited  no emission above $\sim$0.8~keV,  and was classified
as  a so-called  `supersoft source'. 
The X-ray luminosity of this source is one order of magnitude lower than 
the luminosities of the sources S1 and S3. 
This source was fitted by Ott et al. (2003) by using  a black body law. 
Roughly half  of  the supersoft
sources  with optical  counterparts are  yet to  be identified  with a
known type of object (e.g.  Di~Stefano \& Kong 2003).
Unfortunately, we did not detect any radio counterpart associated with this
source.

\section{Summary}
The radio observations presented in this paper found  2 of the 6 discrete sources
detected in X-ray by the Chandra observatory. 
These observations resolved for the first time the SNR 
detected in radio several decades ago.
The  compact radio source with a diameter of about 
0.5\arcsec\,  coincides with a Chandra point source
which also shows characteristics  typical of a SNR. 
The SFR of NGC~3077 based on the size of the detected SNR 
of 0.28 \Msolar yr$^{-1}$ is equal to values derived by 
continuum mm observations and the SFR given by the FIR both 
extinction free tracers of the current SFR. 
The size of the  SNR is about 2 times larger 
than the size of the largest SNR detected in M~82 indicating that the 
star forming event  in NGC~3077 is older than the one in M~82. 
The other detected source with the characteristics of a 
compact HII region, coincides with the X-ray source S3, 
an X-ray binary system. We estimate a flux density of 747 $\mu$Jy for this
source. Assuming that all this energy has a thermal origin 
we estimate that only a few massive stars are necessary to 
ionize the observed nebula. A massive and young stellar cluster observed by the Hubble
Space Telescope coincides with the position of both the S3 X-ray source and the HII region.


\section{Acknowledgments}  
MERLIN is a national facility operated by the University of Manchester 
at Jodrell Bank Observatory on behalf of PPARC. 
I gratefully acknowledges the advice and 
technical support given by Peter Thomasson, Anita Richards 
and other members of the Jodrell Bank Observatory.
I also thank  Elena Terlevich, Gillermo Tenorio-Tagle, Roberto Terlevich,  
Divakara Mayya, Paul O'Neill and Antonio Garc\'{\i}a Barreto for useful discussions.
An extensive report from an anonymous referee greatly improved the final 
version of the paper.

\bsp  
\label{lastpage}  

\begin{thebibliography}{99}  


\bibitem[\protect\citeauthoryear{Aretxaga et
al.}{1999}]{1999ApJ...519L.123A} Aretxaga I., Joguet B., Kunth D., Melnick
J., Terlevich R.~J., 1999, ApJ, 519, L123
\bibitem[\protect\citename{Baugh}{1996}]{1996}  
Baugh C.~M., Cole S., Frenk C.~S., 1996, MNRAS, 283, 1361
\bibitem[\protect\citename{00}{00}]{00}  
Benacchio L., Galletta G., 1981, ApJL, 243, 65
\bibitem[\protect\citename{00}{00}]{00}  
Bi H.~G,  Arp H.,  Zimmermann H.~U., 1994,  A$\&$A,   282, 386
\bibitem[\protect\citename{00}{00}]{00}  
B\"oker T.,  et al., 1999, ApJS, 124, 95
\bibitem[\protect\citename{00}{00}]{00}  
Condon J.~J., 1987, ApJS, 65, 485
\bibitem[\protect\citename{00}{00}]{00}  
Condon J.~J., 1992, ARAA, 30, 575 
\bibitem[\protect\citename{00}{00}]{00}  
Di Stefano R., Kong A.K.H., 2003. ApJ, 592, 884
\bibitem[\protect\citeauthoryear{Dyer et al.}{2001}]{2001ApJ...551..439D}
Dyer K.~K., Reynolds S.~P., Borkowski K.~J., Allen G.~E., Petre R., 2001,
ApJ, 551, 439
\bibitem[\protect\citename{00}{00}]{00}  
Fabbiano, G., 1989, ARAA, 27, 87
\bibitem[\protect\citeauthoryear{Green}{2004}]{2004BASI...32..335G} Green
D.~A., 2004, BASI, 32, 335
\bibitem[\protect\citeauthoryear{Harris et al.}{2004}]{00}
Harris J., Calzetti D., Gallagher J.~S., Smith D.~A., Conselice C.~J.,
2004, ApJ, 603, 503
\bibitem[\protect\citename{00}{00}]{00}  
H{\" o}gbom J.~A., 1974, A$\&$AS, 15, 417
\bibitem[\protect\citename{00}{00}]{00}  
Jarrett et al., 2002, The Two Micron All-Sky Survey: Extended Source Images.  
\bibitem[\protect\citename{00}{00}]{00}  
Kinney  A. L., Bohlin  R. C., Calzetti  D., Panagia  N.,    Wyse R. F. G., 1993, ApJS , 86,  5     
\bibitem[\protect\citeauthoryear{Kronberg \&
Sramek}{1992}]{1992xrea.conf..247K} Kronberg P.~P., Sramek R.~A., 1992,
xrea.conf, 247
\bibitem[\protect\citename{00}{00}]{00}  
Kunth D., Legrand  F., Tenorio-Tagle  G., Silich  S., Mas-Hesse~J.~M.,
Cervi\~ no  M., 2002, Ap\&SS, 281, 261    
\bibitem[\protect\citeauthoryear{Leitherer et
al.}{1999}]{1999ApJS..123....3L} Leitherer C., et al., 1999, ApJS, 123, 3
\bibitem[\protect\citename{00}{00}]{00}  
Martin  C.L., 1998, ApJ, 506, 222
\bibitem[\protect\citename{00}{00}]{00}  
Mas-Hesse J.~M., Kunth D., 1991, A\&AS, 88, 399
\bibitem[\protect\citename{00}{00}]{00}  
McDonald A.~R., Muxlow  T.~W.~B., Wills K.~A., Pedlar A.,  Beswick R.~J., 2002, MNRAS, 334, 912
\bibitem[\protect\citename{00}{00}]{00}  
Meier D.~S., Turner J.~L.,  Beck S.~C., 2001, Astron. J., 122, 1770    
\bibitem[\protect\citename{00}{00}]{00}  
Miley G.,1980, ARAA, 18, 165
\bibitem[\protect\citename{00}{00}]{00}  
Moshir et al. 1990, The IRAS  Faint Source Catalog. 
\bibitem[\protect\citename{00}{00}]{00}  
Muxlow T.W.B., Pedlar A., Wilkinson P.N., Axon D.J., Sanders E.M.,  de Bruyn
A.G., 1994, MNRAS, 266, 455
\bibitem[\protect\citename{00}{00}]{00}  
Niklas S., Klein U., Braine J., Wielebinski R., 1995, A$\&$AS  , 114, 21 
\bibitem[\protect\citename{00}{00}]{00}  
Ott J., Martin C.L.,  Walter F., 2003, ApJ, 594, 776 
\bibitem[\protect\citeauthoryear{Ott, Walter, \&
Brinks}{2005}]{2005MNRAS.358.1453O} Ott J., Walter F., Brinks E., 2005,
MNRAS, 358, 1453
\bibitem[\protect\citeauthoryear{Pannuti et
al.}{2002}]{2002ApJ...565..966P} Pannuti T.~G., Duric N., Lacey C.~K.,
Ferguson A.~M.~N., Magnor M.~A., Mendelowitz C., 2002, ApJ, 565, 966
\bibitem[\protect\citename{00}{00}]{00}  
P\'erez--Torres M.A., Alberdi A., Marcaide J.M., 2004, 7th European VLBI
Network.
\bibitem[\protect\citename{00}{00}]{00}  
Prestwich A. H. et al. 2003, ApJ,  595, 719
\bibitem[\protect\citename{00}{00}]{00}  
Rosa-Gonz\'alez D.,  Terlevich E.,  Terlevich R., 2002. MNRAS, 332, 283 
\bibitem[\protect\citename{00}{00}]{00} 
Seward F.,   Slane P.,  Smith R., and Gaetz T., 2005,  Chandra SN Catalog\footnote{
avaliable electronically at: http://snrcat.cfa.harvard.edu/index.html}
\bibitem[\protect\citeauthoryear{Silich \&
Tenorio-Tagle}{2001}]{2001ApJ...552...91S} Silich S., Tenorio-Tagle G.,
2001, ApJ, 552, 91
\bibitem[\protect\citename{00}{00}]{00}  
Silich S., Tenorio-Tagle G., Mu\~ noz-Tu\~n\' on C., Cairos L.~M., 2002, Astron. J.  , 123, 2438
\bibitem[\protect\citeauthoryear{Tenorio-Tagle et
al.}{1991}]{1991MNRAS.251..318T} Tenorio-Tagle G., Rozyczka M., Franco J.,
Bodenheimer P., 1991, MNRAS, 251, 318
\bibitem[\protect\citeauthoryear{Tenorio-Tagle et
al.}{1990}]{1990MNRAS.244..563T} Tenorio-Tagle G., Bodenheimer P., Franco
J., Rozyczka M., 1990, MNRAS, 244, 563
\bibitem[\protect\citename{00}{00}]{00}  
Tammann G.~A., Sandage A., 1968, ApJ, 151, 825 
\bibitem[\protect\citename{00}{00}]{00}  
Thronson H.~A., Wilton C., Ksir A., 1991, MNRAS, 252, 543  
\bibitem[\protect\citename{00}{00}]{00}  
Walter F., Weiss A., Martin C., Scoville N., 2002, Astron. J., 123, 225 
\bibitem[\protect\citename{00}{00}]{00}  
Walter F., Kerp J., Duric N., Brinks E.,  Klein U., 1998, ApJL,   502, 143 
\end{thebibliography}
\end{document}